\documentclass{revtex4}

\usepackage{amsthm,amssymb,amsmath}				
\usepackage{graphicx,comment}
\usepackage{float}   
 \usepackage{xcolor}

\begin{document}

\title{ Klein-Gordon oscillators in traversable wormhole rainbow gravity spacetime: Conditional exact solvability via a throat radius and oscillator frequency correlation}
\author{Omar Mustafa}
\email{omar.mustafa@emu.edu.tr}
\affiliation{Department of Physics, Eastern Mediterranean University, 99628, Famagusta, North Cyprus, Mersin -10, Türkiye.}
\author{Abdullah Guvendi}
\email{abdullah.guvendi@erzurum.edu.tr}
\affiliation{Department of Basic Sciences, Erzurum Technical University, 25050, Erzurum, Türkiye.}

\begin{abstract}
\textbf{Abstract:}\ In this study, we discuss an analytical solution for a set of the Klein-Gordon (KG) oscillators' energies through a correlation between the frequency of the KG-oscillators and the traversable wormhole (TWH) throat radius. Under such restricted parametric correlation (hence the notion of conditionally exact solvability is unavoidable in the process), we report the effects of throat radius, rainbow parameter, disclination parameter, and oscillator frequency on the spectroscopic structure of a vast number of $\left( n,m\right) $-states (the radial and magnetic quantum numbers, respectively). In the process, we only use two loop quantum gravity motivated rainbow functions pairs. The only rainbow functions that clearly and reliably fully adhere to the rainbow gravity model and secure Planck energy $E_p$ as the maximum possible energy for particles and anti-particles alike. Near the asymptotically flat upper and lower universes connected by the TWH, i.e., for the throat radius  $r_{0 }>>1 $, the energies tend to cluster around the rest mass energies, i.e.,  $|E_{\pm }|\sim  m_{0 }$.  Whereas, for $r_{0 }<<1$, the energies tend to approach $|E_{\pm }| \leq E_{P}$.


\end{abstract}

\maketitle

\section{Introduction}

In the last few decades, wormholes, as solutions to Einstein's field
equation, have attracted research attention. The undoubtable significance of
the effects of gravitational fields on the quantum mechanical structures,
during the early universe, suggests that wormholes (also known as
Einstein-Rosen-bridges (ER) \cite{R1.1}), play a crucial role in cosmology 
\cite{R1.1,R1,R2,R3,R4,R5}. The current accumulated knowledge of physics
does not rule out some intriguing ideas like the feasible existence of
advanced civilizations' capability to build wormholes for the sake of
interstellar and/or time travels \cite{R1,R5.1}. Wormholes in
(2+1)-dimensional spacetime are not only pedagogically interesting but they
are also instructive for ideas related with quantization of gravitational
fields \cite{R5.2}. Such a (2+1)-dimensional spacetime allows comprehensive analysis
of some interesting physical properties of the wormholes spacetime, like
Cauchy horizon properties \cite{R5.1}, wormholes stability \cite{R5.3}, testing wormhole solutions in extended gravity \cite{R5.31,R5.32,R5.33}, in Rastall-Rainbow gravity \cite{R5.34,R5.35,R5.36} and in 4D EGB gravity \cite{R5.37}, etc. In quantum gravity, nevertheless, the harmonization of quantum mechanics with general relativity, has manifestly inspired intricate studies
on quantum particles and/or antiparticles within the wormholes spacetime.
Such intricate studies have shown that while the gravitational force fields
introduce intense curvature, in the vicinity of wormholes spacetime, they
may influence the dynamics of quantum particles \cite%
{R5.4,R5.5,R5.6,R5.7,R5.8,R5.9,R5.10,R5.11}, and may yield phenomena of time dilation and particle's trajectory distortion \cite{R5.3}, to mention a few.

Unlike black holes, wormholes connect a region of the universe to another
universe (called upper universe and lower universe \cite{R5}) or to a distant part of the same universe without reaching a
non-removable singularity \cite{R1}. Their geometry possesses a non-zero
minimum of the radial distance $r$ so that $r_{\min }=r_{0 }\neq 0$ is
the so-called throat radius of the wormhole. Yet, the metric of the wormhole
spacetime indulges a singularity at the radial coordinate $r_{\min
}=r_{0 }$. Upon imposing Morris-Thorne  traversability conditions, one would require that the (2+1)-dimensional traversable wormhole (TWH)
spacetime metric be radially symmetric and static \cite{R5,R6}. Under such
traversability condition, the\ (2+1)-TWH metric, in $c=\hbar =G=1$ units,
reads \cite{R1,R5,R6,R7}
\begin{equation}
ds^{2}=-e^{2\Phi \left( r\right) }\,dt^{2}+\,\frac{dr^{2}}{1-b\left(
r\right) /r}+r^{2}\,d\varphi ^{2},  \label{1}
\end{equation}%
where $\Phi \left( r\right) $ and $b\left( r\right) $ are arbitrary radial functions, and are the redshift function and the form/shape function, respectively. In the current methodical proposal, we shall focus on a particularly interesting redshift function is $\Phi ^{\prime
}\left( r\right) =0$ (i.e., constant redshift function). Hence without any loss of generality one may use $\Phi \left( r\right) =0$ . Such a redshift function setting yields a significantly simplistic field equation and provides intriguing solutions, which differ from their general relativistic (GR) counterparts. On the other hand, moreover, we consider the form/shape function $b\left( r\right) =r_{0 }^{2}/r$, which is a case that corresponds to negative energy density in GR. Under functional settings, the metric for TWH embedded with disclination ($\propto \alpha$) \cite{wh1,wh2}) would read
\begin{equation}
ds^{2}=-dt^{2}+dx^{2}+R\left( x\right) d\varphi ^{2};\;R\left( x\right)
=\alpha ^{2}\left( x^{2}+r_{0 }^{2}\right) ,  \label{2}
\end{equation}%
where $x^{2}=r^{2}-r_{0 }^{2}$, $r_{0 }$ is the wormhole throat radius, $0<\alpha ^{2}=1-8\pi \eta ^{2}\leq 1$, $\alpha $ is the disclination
parameter related with the deficit angle, $\eta $ stands for the energy density, and $G$ is the gravitational constant. 

Rainbow gravity (RG), on the other hand, suggests that the energy of the probe particle affects the spacetime background, at the ultra-high energy regime, and the spacetime metric (TWH-metric, in our case) takes the energy-dependent form (c.f., e.g., \cite{R8,R9,R10,R11,R12,R13,R14})%
\begin{equation}
ds^{2}=-\frac{1}{g_{_{0 }}\left( u\right) ^{2}}dt^{2}+\frac{1}{%
g_{_{1}}\left( u\right) ^{2}}\left( dx^{2}+R\left( x\right) d\varphi
^{2}\right) ,  \label{3}
\end{equation}%
where $g_{_{k}}\left( u\right) $, $k=0,1$, are the rainbow functions that satisfy the condition: $\lim\limits_{u\rightarrow 0}g_{_{k}}\left( u\right) =1
$, and $u=|E|/E_{p}$ is a fine tuned rainbow gravity parameter (to allow rainbow gravity to affect relativistic particles and anti-particles alike, see e.g. \cite{R15}). The selection of rainbow functions is guided by both phenomenological and theoretical considerations. These functions are chosen because they align with the principles of non-commutative geometry and loop quantum gravity.Such rainbow functions play a crucial role that secures the Planck's energy $E_p$ as a threshold between quantum and classical descriptions and is yet another invariant along side with the speed of light.
In the current study, we shall be interested in two loop quantum
gravity motivated pairs $g_{_{0}}(u)=1$, $g_{_{1}}\left( u\right) =\sqrt{%
1-\epsilon u^{2}}$, and $g_{_{0}}(u)=1$, $g_{_{1}}\left( u\right) =\sqrt{%
1-\epsilon u}$, where $\epsilon \geq 1$ is the rainbow gravity parameter \cite{R15,R15.1,R16,R17} (notice $\epsilon =0$ is used to switch off rainbow gravity). Our interest in the loop quantum gravity rainbow functions steams from the fact that the particle's and anti-particle's energies, under such rainbow functions, completely comply with the rainbow gravity model and secure Planck energy $E_p$ as the maximum energy for particles and anti-particles. Whereas, the rainbow function pairs, i.e., $g_{_0}(u)=g_{_1}(u)= (1-\epsilon u)^{-1},$ used in resolving the horizon problem and those obtained from the spectra of gamma-ray bursts, i.e, $g_{_0}(u)= (e^{\epsilon u}-1)$ and $g_{_1}(u)=1$,  have only partially complied with the intended rainbow gravity effect (for more details on this issue the reader is advised to refer to \cite{R15,R15.1}). 

Most approaches to quantum gravity (e.g., string field theory \cite{R18}, loop quantum gravity \cite{R19}, and non-commutative geometry \cite{R20}) had a common suggestion that modifies the standard relativistic energy-momentum dispersion relation in the ultraviolet regime into%
\begin{equation}
E^{2}g_{_{0}}\left( u\right) ^{2}-p^{2}g_{_{1}}\left( u\right)
^{2}=m^{2}.  \label{3.1}
\end{equation}%
Where, in this case, the condition $\lim\limits_{u\rightarrow
0}g_{_{k}}\left( u\right) =1$ retrieves the standard energy-momentum
dispersion relation in the infrared regime. Such modified dispersion relations are natural byproducts of the doubly/deformed special relativity (DSR) \cite{R21,R22,R23,R24,R25}. DSR is an extension to special relativity
that introduces yet another invariant energy scale (the Planck's energy $%
E_{p}$), alongside with the invariance of the speed of light. A generalization of DSR that includes curvature is provided by the doubly general relativity \cite{R26}, where the spacetime metric depends on the energy of the probe particle and forms the so-called rainbow gravity (RG). Studies on the RG effects are carried out, for example, on the thermal properties of black holes \cite{R27,R28,R29,R30,R31,R32,R33}, on the ultra high-energy cosmic rays \cite{R28,R34,R35}, on massive scalar field in RG-Schwarzschild metric \cite{R36}, five-dimensional Yang--Mills black holes in massive RG \cite{R37}, to mention a few.

Recent studies on the dynamics of relativistic Klein-Gordon (KG) and Dirac particles were carried out using different spacetimes in rainbow gravity backgrounds. For example, in a cosmic string spacetime background in rainbow gravity, Landau levels via Schr\"{o}dinger and KG equations were reported by Bezerra et al. \cite{R38}, Dirac oscillators by Bakke and Mota \cite{R39,R40}, quantum dynamics of photon by Sogut et al.  \cite{R41}, KG-particles in a topologically trivial G\"{o}del-type spacetime in rainbow gravity by Kangal et al. \cite{R42}, KG-Coulomb particles in cosmic string rainbow gravity spacetime \cite{R14} and massless KG-oscillators in Som-Raychaudhuri cosmic string spacetime in a fine tuned rainbow gravity \cite{R15} by Mustafa. The studies of such systems are motivated by the fact that the determination of their exact, quasi-exact, or even conditionally exact solutions allows one to describe the dynamics of relativistic/non-relativistic particles in the curved spacetime. Which, in turn, facilitates the understanding of the intricate interplay between quantum mechanics and general relativity. These solutions may reveal fascinating influence of the curved spacetime on the spectroscopic structure of relativistic particles that have implications in cosmology, geometrical
theory of topological defect, study of black holes, wormholes, to mention a few areas. 

Motivated by the intriguing effects of the topology of different spacetimes backgrounds on the spectroscopic structures of relativistic and non-relativistic quantum particles, we, in this work, investigate KG-oscillators in TWH-spacetime background in rainbow gravity. In so doing, we provide, in section 2, the mathematical foundation for the corresponding KG-particles, with  $\mathcal{F}_{i}=\left( 0,\mathcal{F}_{x},0\right) $ in the non-minimal coupling form $\tilde{D}^\pm_{i}=\partial_{i}\pm\mathcal{F}_{i}$.  In section 3, we incorporate the KG-oscillator using the assumption that $\mathcal{F}_{x}=\Omega x$ and
discuss one of the conditionally exact solution that allows comprehensive analysis on the effects of 
(i)  the rainbow gravity parameter $\epsilon$, indulged within $\beta=\epsilon/E_p$, (ii) the effect of the throat radius $r_0$ of the TWH,  (iii) the effects of the KG-oscillator frequency $\Omega $, and (iv) the effect of the disclination parameter,  on the spectroscopic structure of KG-oscillators. We use two loop quantum gravity motivated pairs $%
g_{_{0}}(u)=1$, $g_{_{1}}\left( u\right) =\sqrt{1-\epsilon u^{2}}$, and $%
g_{_{0}}(u)=1$, $g_{_{1}}\left( u\right) =\sqrt{1-\epsilon u}$, where $%
\epsilon \geq 1$ is the rainbow gravity parameter \cite{R15,R16,R17}. Our concluding remarks are given in section 4.

\section{KG-particles in TWH-spacetime in rainbow gravity}

The covariant and contravariant metric tensors of the TWH-spacetime in rainbow gravity (\ref{3}), respectively, read
\begin{equation}
g_{ij}=diag\left( -\frac{1}{g_{_{0 }}\left( y\right) ^{2}},\frac{1}{%
g_{_{1}}\left( y\right) ^{2}},\frac{R\left( x\right) }{g_{_{1}}\left(
y\right) ^{2}}\right) ;\;i,j=t,x,\varphi \Longrightarrow \det \left(
g_{ij}\right) =g=-\frac{R\left( x\right) }{g_{_{0 }}\left( y\right)
^{2}g_{_{1}}\left( y\right) ^{4}},  \label{4}
\end{equation}
and 
\begin{equation}
g^{ij}=diag\left( -g_{_{0 }}\left( y\right) ^{2},\,g_{_{1}}\left(
y\right) ^{2},\frac{g_{_{1}}\left( y\right) ^{2}}{R\left( x\right) }\right). 
\label{5}
\end{equation}
The KG-equation would then read 
\begin{equation}
\left( \frac{1}{\sqrt{-g}}\tilde{D}^+_{i}\sqrt{-g}g^{ij}\tilde{D}^-_{j
}\right) \,\Psi \left( t,x,\varphi \right) =m_{0 }^{2}\,\Psi \left(
t,x,\varphi \right) ,  \label{6}
\end{equation}%
where $\tilde{D}^\pm_{i}=\partial_{i}\pm\mathcal{F}_{i}$ is in a non-minimal coupling form with $\mathcal{F}_{i}$ $\in 
\mathbb{R}
$, and $m_{0 }$ is the rest mass energy (i.e., $%
m_{0 }\equiv m_{0 }c^{2}$). Moreover, we use $\mathcal{F}_{i}=\left(
0,\mathcal{F}_{x},0\right) $ to incorporate the KG-oscillators in the TWH-spacetime in rainbow gravity. Consequently, the KG-equation (\ref{6}) would read
\begin{equation}
\left( \frac{1}{\sqrt{-g}}\left( \partial _{i}+\mathcal{F}%
_{i}\right) \sqrt{-g}g^{ij}\left( \partial _{j}-\mathcal{F}%
_{j}\right) \right) \,\Psi \left( t,x,\varphi \right) =m_{0 }^{2}\,\Psi
\left( t,x,\varphi \right) ,  \label{6.1}
\end{equation}
which, in turn, yields
\begin{equation}
\left\{ -g_{_{0 }}\left( y\right) ^{2}\partial _{t}^{2}+g_{_{1}}\left(
y\right) ^{2}\left[ \partial _{x}^{2}+\frac{x}{x^{2}+r_{0 }^{2}}\partial
_{x}-\mathcal{M}\left( x\right) +\frac{ \partial _{\varphi
} ^{2}}{R\left( x\right) }\right] \right\} \Psi \left(
t,x,\varphi \right) =m_{0 }^{2}\,\Psi \left( t,x,\varphi \right) ,
\label{7}
\end{equation}
where
\begin{equation}
\mathcal{M}\left( x\right) =\frac{x}{x^{2}+r_{0 }^{2}}\mathcal{F}_{x}+%
\mathcal{F}_{x}^{\prime }+\mathcal{F}_{x}^{2}.  \label{8}
\end{equation}
We now use the usual substitution 
\begin{equation}
\Psi \left( t,x,\varphi \right) =\Psi \left( t,r,\varphi \right) =\exp
\left( -iEt+im\varphi \right) \,\psi \left( x\right) ,  \label{8.1}
\end{equation}%
where $m=0,\pm 1.\pm 2,\cdots $ is the magnetic quantum number, to obtain
\begin{equation}
\left\{ \partial _{x}^{2}+\frac{x}{x^{2}+r_{0 }^{2}}\,\partial _{x}-%
\mathcal{M}\left( x\right) -\frac{ m^{2}}{\alpha
^{2}\left( x^{2}+r_{0 }^{2}\right) }+\mathcal{E}\right\} \psi \left(
x\right) =0,  \label{8.2}
\end{equation}
where
\begin{equation*}
\mathcal{E}=\frac{E^{2}g_{_{0 }}\left( y\right) ^{2}-m_{0 }^{2}}{%
g_{_{1}}\left( y\right) ^{2}}.
\end{equation*}%
This equation represents a KG-particle in the TWH-spacetime under rainbow gravity settings.

\section{KG-oscillators in TWH-spacetime in rainbow gravity}

The substitution of $\mathcal{F}_{x}=\Omega x$ in (\ref{8.2}) implies that
\begin{equation}
\left\{ \partial _{x}^{2}+\frac{x}{x^{2}+r_{0 }^{2}}\,\partial
_{x}-\Omega ^{2}x^{2}-\frac{\Omega \,x^{2}}{x^{2}+r_{0 }^{2}}-\frac{
 m ^{2}}{\alpha ^{2}\left( x^{2}+r_{0
}^{2}\right) }+\mathcal{\tilde{E}}\right\} \psi \left( x\right) =0,
\label{9}
\end{equation}
where
\begin{equation}
\mathcal{\tilde{E}}=\frac{E^{2}g_{_{0 }}\left( y\right) ^{2}-m_{0
}^{2}}{g_{_{1}}\left( y\right) ^{2}}-\Omega .  \label{9.1}
\end{equation}%
This equation would take the form%
\begin{equation}
\left\{ \partial _{x}^{2}+\frac{x}{x^{2}+r_{0 }^{2}}\,\partial
_{x}-\Omega ^{2}x^{2}-\frac{\Omega \,x^{2}}{x^{2}+r_{0 }^{2}}-\frac{%
\tilde{m}^{2}}{\left( x^{2}+r_{0 }^{2}\right) }+\mathcal{\tilde{E}}%
\right\} \psi \left( x\right) =0, \label{10}
\end{equation}
where $\tilde{m}=m/\alpha $.  One would notice that the fourth term of this equation was missed in Eq.(6) of \cite{R5.11}, as a consequence of the mistake made in their Eq. (5) (namely, the last term, first line, of their (5) $M\omega a^2/(x^2+a^2)$ should be replaced by $-M\omega x^2/(x^2+a^2)$, our $\Omega=M\omega$ in \cite{R5.11} ).  This would change all the  results reported therein. That is, equation (\ref{10}) is known to admit a solution in the form of confluent Heun function so that
\begin{equation}
\psi \left( x\right) =C\,\exp (-\frac{\Omega x^{2}}{2})\,H_{C}\left( \Omega
r_{0 }^{2},-\frac{1}{2},-\frac{1}{2},-\frac{r_{0 }^{2}\left( 
\mathcal{\tilde{E}}-\Omega \right) }{4},\frac{r_{0 }^{2}\mathcal{\tilde{E%
}-}\tilde{m}^{2}}{4}+\frac{3}{8},-\frac{x^{2}}{r_{0 }^{2}}\right). 
\label{11}
\end{equation}
At this point, one should notice that for $\,H_{C}\left( \grave{\alpha},
\grave{\beta},\grave{\gamma},\grave{\delta},\grave{\eta},z\right) $ we have
\begin{equation}
\grave{\delta}=-\grave{\alpha}\left( n+\frac{1}{2}\left[ \grave{\beta}+
\grave{\gamma}+2\right] \right) \Longrightarrow -\frac{r_{0 }^{2}\left( 
\mathcal{\tilde{E}}-\Omega \right) }{4}=-\Omega r_{0 }^{2}\left( n+\frac{
1}{2}\right) \Longrightarrow \mathcal{\tilde{E}}=4\Omega \left( n+\frac{3}{4}
\right),   \label{12}
\end{equation}
with $n\geq 0$ so that the confluent Heun function is truncated into a polynomial of order $n+1$ \cite{RR1,RR2,RR2.1}. Yet, one may use
\begin{equation}
\psi \left( x\right) =\exp \left( -\frac{\Omega \,x^{2}}{2}\right) \,\phi
\left( x\right),   \label{13}
\end{equation}%
in (\ref{10}) to obtain
\begin{equation}
\left( x^{2}+r_{0 }^{2}\right) \,\phi ^{\prime \prime }\left( x\right) +
\left[ \left( 1-2\Omega r_{0 }^{2}\right) \,x-2\Omega \,x^{3}\right]
\,\,\phi ^{\prime }\left( x\right) +\left[ \left( \mathcal{\tilde{E}}
-3\Omega \right) x^{2}+\left( \mathcal{\tilde{E}}-\Omega \right) r_{0
}^{2}-\tilde{m}^{2}\right] \,\,\phi \left( x\right) =0.  \label{14}
\end{equation}
Let us use a change of variable in the form of $y=x^{2}$ to obtain
\begin{equation}
y\left( y+r_{0 }^{2}\right) \,\phi ^{\prime \prime }\left( y\right) +
\left[ \left( 1-\Omega r_{0 }^{2}\right) y-\Omega \,y^{2}+\frac{r_{0
}^{2}}{2}\right] \phi ^{\prime }\left( y\right) +\left[ P_{1}\,y+P_{2}\right]
\,\phi \left( y\right) =0,  \label{15}
\end{equation}
where,
\begin{equation}
P_{1}=\frac{\mathcal{\tilde{E}}}{4}-\frac{3}{4}\Omega \text{\thinspace }%
\,,\;P_{2}=\frac{\left( \mathcal{\tilde{E}}-\Omega \right) r_{0 }^{2}}{4}%
-\frac{\tilde{m}^{2}}{4}.  \label{16}
\end{equation}%
A power series expansion in the form of
\begin{equation}
\phi \left( y\right) =y^{\sigma }\sum\limits_{j=0}^{\infty }C_{j}\,y^{j},
\label{17}
\end{equation}
in (\ref{15}) to yield
\begin{gather}
\sum\limits_{j=0}^{\infty }C_{j}\left[ P_{1}-\Omega \left( j+\sigma \right) 
\right] \,y^{j+\sigma +1}\,+\sum\limits_{j=-1}^{\infty }C_{j+1}\left[
P_{2}+\left( j+\sigma +1\right) \left( j+\sigma +1-\Omega r_{0
}^{2}\right) \right] \,y^{j+\sigma +1}  \notag \\
+\sum\limits_{j=-2}^{\infty }C_{j+2}\left[ r_{0 }^{2}\left( j+\sigma
+2\right) \left( j+\sigma +\frac{3}{2}\right) \right] \,y^{j+\sigma +1}=0.
\label{18}
\end{gather}
This equation would in turn imply that
\begin{gather}
\sum\limits_{j=0}^{\infty }\left\{ C_{j}\left[ P_{1}-\Omega \left( j+\sigma
\right) \right] +C_{j+1}\left[ P_{2}+\left( j+\sigma +1\right) \left(
j+\sigma +1-\Omega r_{0 }^{2}\right) \right] \right.   \notag \\
\left. +C_{j+2}\left[ r_{0 }^{2}\left( j+\sigma +2\right) \left(
j+\sigma +\frac{3}{2}\right) \right] \right\} \,y^{j+\sigma +1}=0,  \label{19}
\end{gather}
where 
\begin{equation}
\sigma \left( \sigma -\frac{1}{2}\right) =0\Longrightarrow \sigma
=0,\;\sigma =\frac{1}{2},  \label{20}
\end{equation}%
and 
\begin{equation}
C_{1}\left[ r_{0 }^{2}\left( \sigma +1\right) \left( \sigma +\frac{1}{2}%
\right) \right] +C_{0}\left[ P_{2}+\sigma \left( \sigma -\Omega r_{0
}^{2}\right) \right] =0.  \label{21}
\end{equation}

Obviously, equation (\ref{20}) mandates that $\sigma =1/2$ is quantum
mechanically not interesting. One should observe that $y^{\sigma }$ in (\ref%
{17}) would result, with $\sigma =1/2$, that $\sqrt{y}=x=\sqrt{%
r^{2}-r_{0 }^{2}}$. Consequently, $x=0$ corresponds to $r=r_{0 }$
(at the wormhole throat) and the wave function should not vanish at that
point (unless a hard wall is assumed to exist at $r=r_{0 }$ (which is
not the case under consideration in the current proposal). We therefore
adopt $\sigma =0$ and use (\ref{21}) to obtain
\begin{equation}
C_{1}=-\frac{2P_{2}}{r_{0 }^{2}}C_{0}  \label{22}
\end{equation}
where $C_{0}=1$ is to be used hereinafter. Under such settings, one obtains
\begin{equation}
C_{j}\left[ P_{1}-\Omega j\right] \,\,+C_{j+1}\left[ P_{2}-\left( j+1\right)
\left( j+1-\Omega r_{0 }^{2}\right) \right] \,+C_{j+2}\left[ r_{0
}^{2}\left( j+2\right) \left( j+\frac{3}{2}\right) \right] =0,  \label{23}
\end{equation}
to yield
\begin{equation}
C_{j+2}=\frac{C_{j}\left[ \Omega j-P_{1}\right] +C_{j+1}\left[ \left(
j+1\right) \left( j+1-\Omega r_{0 }^{2}\right) -P_{2}\right] }{r_{0
}^{2}\left( j+2\right) \left( j+\frac{3}{2}\right) };\;j=0,1,2,\cdots.
\label{24}
\end{equation}
Consequently, we get for $j=0$
\begin{equation}
C_{2}=\frac{C_{0}\left[ -P_{1}\right] +C_{1}\left[ \left( 1-\Omega r_{0
}^{2}\right) -P_{2}\right] }{3r_{0 }^{2}},  \label{24.1}
\end{equation}
for $j=1$
\begin{equation}
C_{3}=\frac{C_{1}\left[ \Omega -P_{1}\right] +C_{2}\left[ \left( 2\right)
\left( 2+1-\Omega r_{0 }^{2}\right) -P_{2}\right] }{\frac{15}{2}r_{0
}^{2}},  \label{24.2}
\end{equation}
and so on. However, to secure finiteness and square integrability of the corresponding wave function, we need to truncate the power series into a polynomial by requiring that for $\forall j=n$ we have $C_{n+2}=0$. This
assumption would yield a polynomial of order $n+1\geq 1$ \cite{RR1,RR2,RR2.1,RR2.2,RR2.3,RR2.4,RR2.5}. Moreover, we may seek the condition that for $\forall j=n$ and $
C_{n+1}\neq 0$ we suggest that 
\begin{equation}
P_{2}=\left( n+1\right) \left( n+1-\Omega r_{0 }^{2}\right)
\Longrightarrow \frac{\mathcal{\tilde{E}}r_{0 }^{2}}{4}=\frac{\tilde{m}
^{2}}{4}+\left( n+1\right) ^{2}-\Omega r_{0 }^{2}\left( n+\frac{3}{4}
\right).  \label{25}
\end{equation}
This would, in turn, facilitate the so called \textit{conditional exact solvability} of the system at hand and consequently allows us to take
\begin{equation}
P_{1}-\Omega n=0\Longrightarrow \frac{\mathcal{\tilde{E}}}{4}-\frac{3}{4}%
\Omega \text{\thinspace }=\Omega n\Longrightarrow \mathcal{\tilde{E}}
=4\Omega \left( n+\frac{3}{4}\right),  \label{26}
\end{equation}
since $C_{n}\neq 0$. It is clear that the result in (\ref{26}) is in exact accord with that in (\ref{12}). Yet, the condition in (\ref{25}) along with (\ref{26}) would imply that
\begin{equation}
\Omega r_{0 }^{2}=\frac{4\alpha ^{2}\left( n+1\right) ^{2}+m^{2}}{
4\alpha ^{2}\left( 2n+3/2\right)}.  \label{27}
\end{equation}
Notably, this result identifies a correlation between the throat radius $
r_{0 }$ and the frequency of the KG-oscillator $\Omega $ at a specific $
\alpha $ value. Such a correlation suggests that KG-oscillator frequency $\Omega $ is inversely proportional with square the throat radius $r_0$. This correlation, obviously, not only inspires the truncation condition of the confluent Heun function in (\ref{12}) into a polynomial of order $n+1$, but also offers a \textit{conditionally exactly solvable} system (i.e., it only describes a restricted set of the oscillator frequencies, $\Omega^{'}s\rightarrow \Omega_{n,m} {}^{'}s$,  given by this correlation, and consequently describes the set of energy levels associated with such restricted oscillators' frequencies. Hence the notion of  \textit{conditional exact solvability} is unavoidable in the process ). Using (\ref{9.1}), we obtain
\begin{equation}
\mathcal{\tilde{E}}=4\Omega \left( n+\frac{3}{4}\right) \Longrightarrow
E^{2}g_{_{0 }}\left( u\right) ^{2}=m_{0 }^{2}+4\Omega \left(
n+1\right) \;g_{_{1}}\left( u\right) ^{2}.  \label{28}
\end{equation}%
Finally, the corresponding wave functions are given by 
\begin{equation}
\Psi \left( t,x,\varphi \right) =\exp
\left( -iEt+im\varphi \right) \,\exp \left( -\frac{\Omega \,x^{2}}{2}\right) \,\phi
\left( x\right),  \label{28.1}
\end{equation}%
where $\phi\left( x\right)$ are now the confluent Heun polynomials given by
\begin{equation}
\phi \left( x\right) =\sum\limits_{j=0}^{n+1 }C_{j}\,x^{2j},
\label{28.2}
\end{equation}%
This result is to be solved for some loop quantum gravity motivated rainbow function pairs $g_{_{0}}(u)=1$, $g_{_{1}}\left( u\right) =\sqrt{1-\epsilon u}
$, and $g_{_{0}}(u)=1$, $g_{_{1}}\left( u\right) =\sqrt{1-\epsilon u^{2}}$, with $u=|E|/E_{P}$.

\subsection{Rainbow function pair $g_{_{0}}(u)=1$, $g_{_{1}}\left( u\right) =
\sqrt{1-\epsilon u};\;u=|E|/E_{P}$}

Under such a rainbow function structure in (\ref{28}), one obtains
\begin{equation}
E^{2}=m_{0 }^{2}+4\Omega \left( n+1\right) \left( 1-\epsilon \frac{|E|}{%
E_{P}}\right) \Longleftrightarrow \;E^{2}+4\Omega \beta \left( n+1\right)
|E|-\left[ m_{0 }^{2}+4\Omega \left( n+1\right) \right] =0,  \label{29}
\end{equation}
where $\beta =\epsilon /$ $E_{P}$, and $E=E_{\pm }=\pm |E|$ indulges particle, $E_{+}=+|E|$, and anti-particle, $E_{-}=-|E|$, energies. In this case, one gets two equations from (\ref{29})
\begin{equation}
\;E_{+}^{2}+4\Omega \beta \left( n+1\right) E_{+}-\left[ m_{0
}^{2}+4\Omega \left( n+1\right) \right] =0,  \label{30}
\end{equation}
hence, we get
\begin{equation}
E_{+}=-2\Omega \beta \left( n+1\right) +\sqrt{4\Omega ^{2}\beta ^{2}\left(
n+1\right) ^{2}+m_{0 }^{2}+4\Omega \left( n+1\right) },  \label{31}
\end{equation}%
for particles, and
\begin{equation}
\;E_{-}^{2}-4\Omega \beta \left( n+1\right) E_{-}-\left[ m_{0
}^{2}+4\Omega \left( n+1\right) \right] =0,  \label{32}
\end{equation}
hence,
\begin{equation}
E_{-}=+2\Omega \beta \left( n+1\right) -\sqrt{4\Omega ^{2}\beta ^{2}\left(
n+1\right) ^{2}+m_{0 }^{2}+4\Omega \left( n+1\right) },  \label{33}
\end{equation}
for antiparticles. Nevertheless, one may clearly observe that the energy levels admit symmetry about $E=0$. In our opinion, this should always be the tendency of the energy levels of relativistic particles. 
\begin{figure}[ht!]  
\centering
\includegraphics[width=0.3\textwidth]{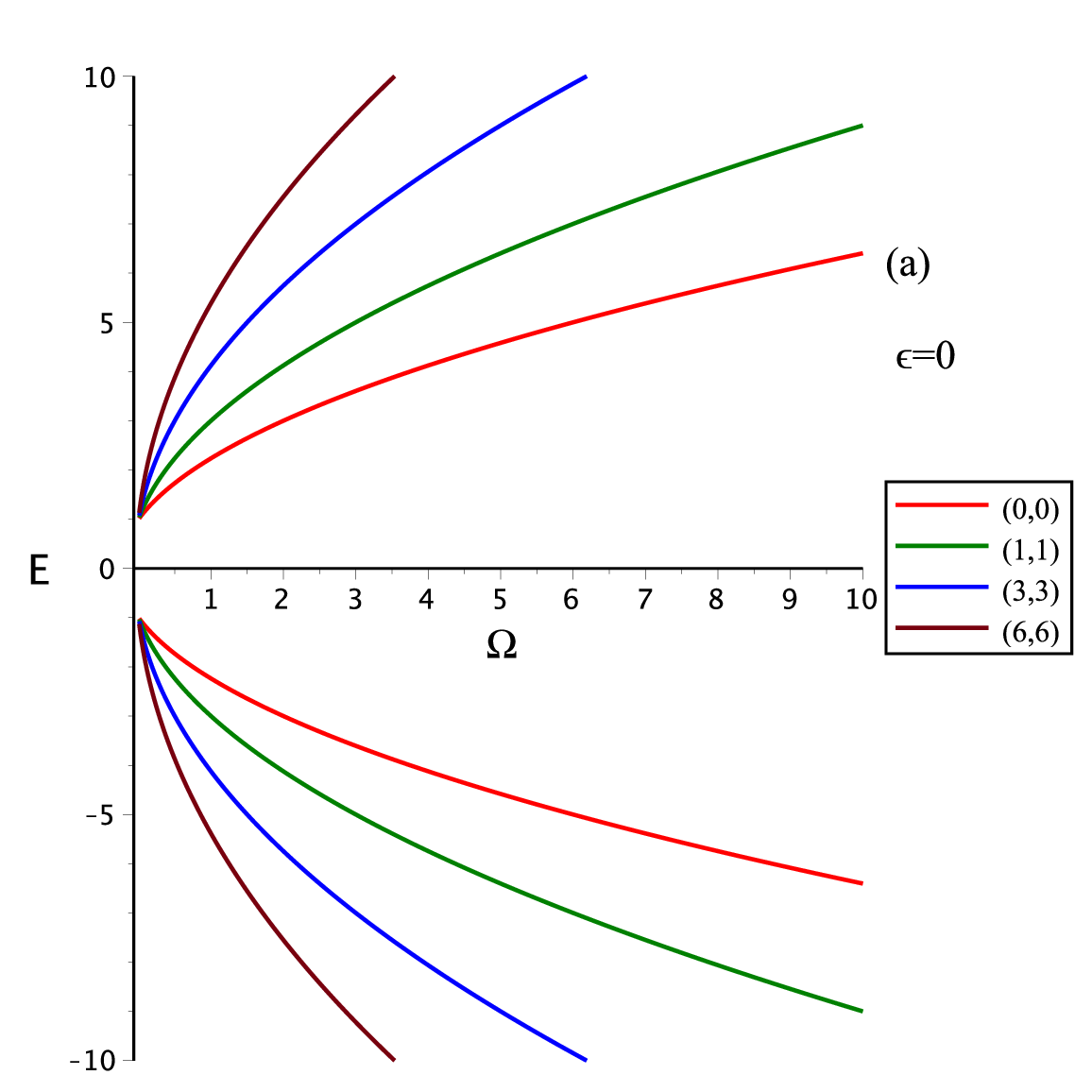}
\includegraphics[width=0.3\textwidth]{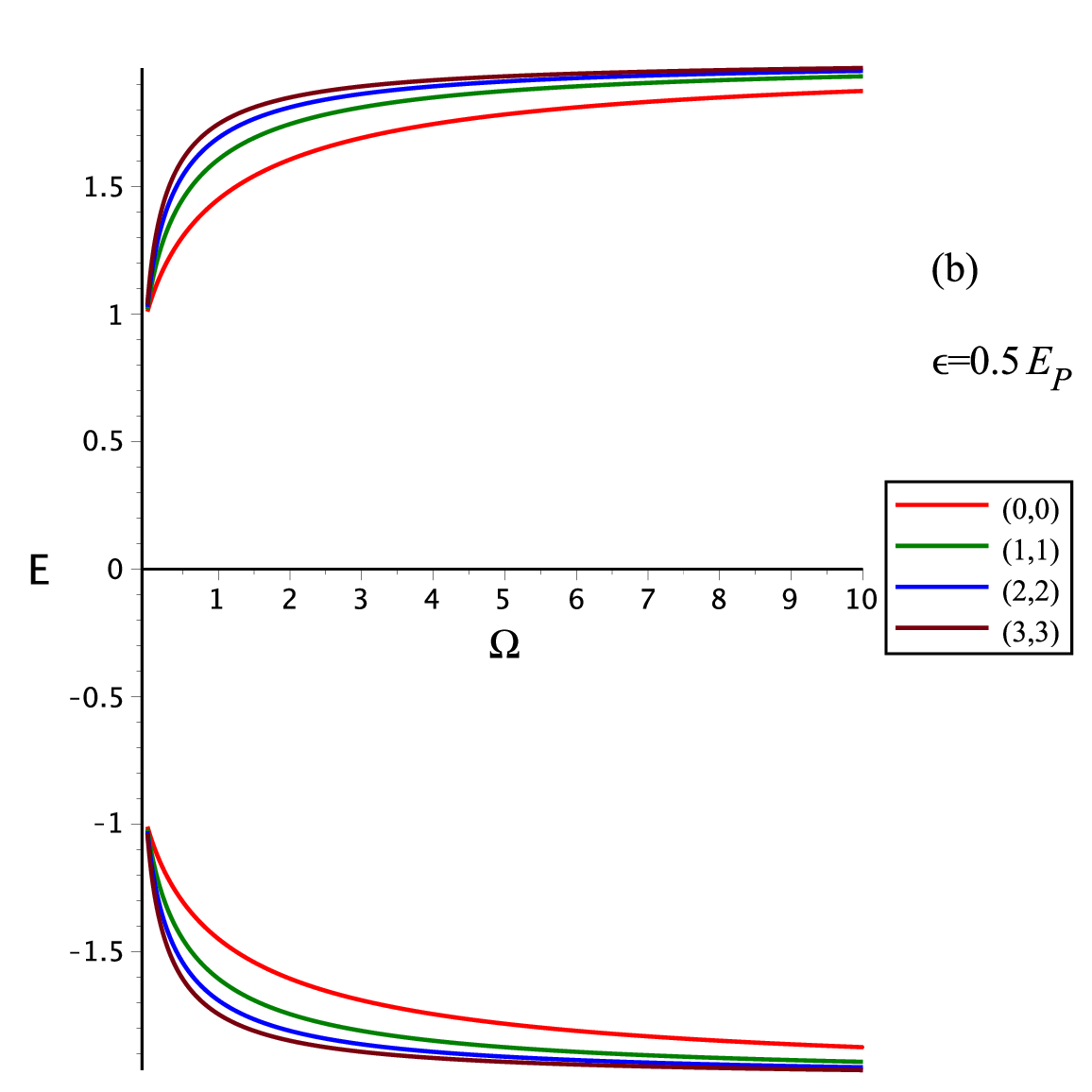}
\includegraphics[width=0.3\textwidth]{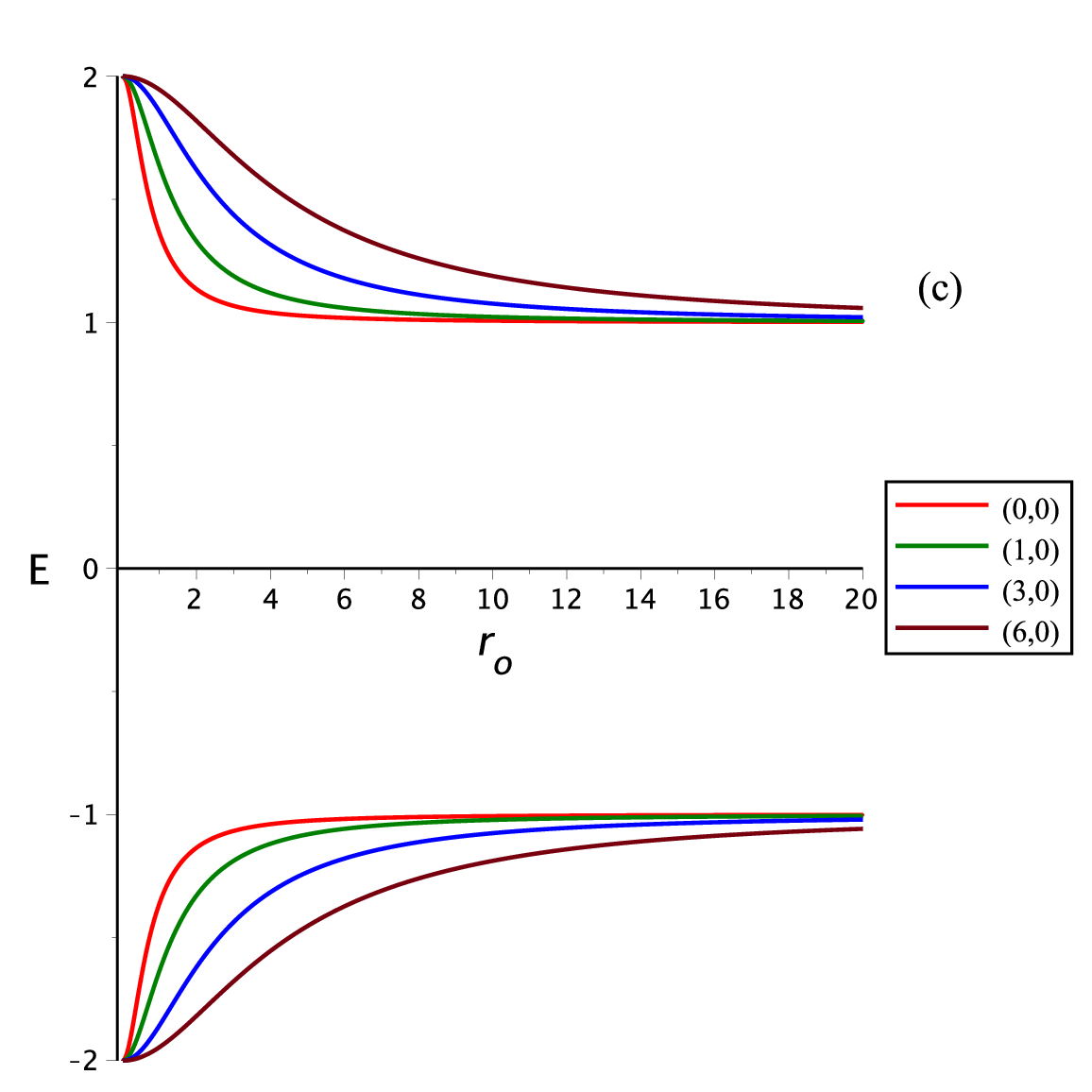}
\includegraphics[width=0.3\textwidth]{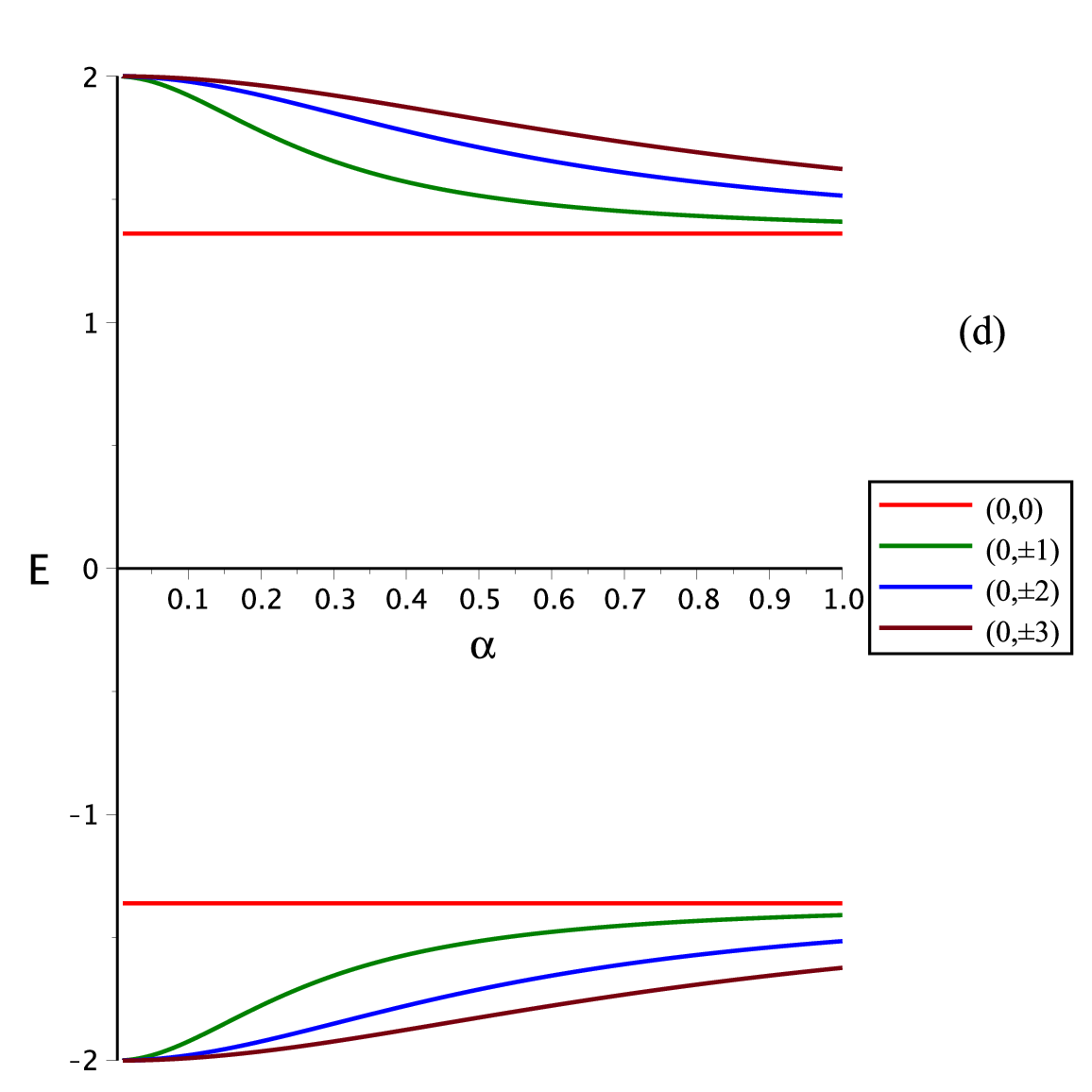}
\includegraphics[width=0.3\textwidth]{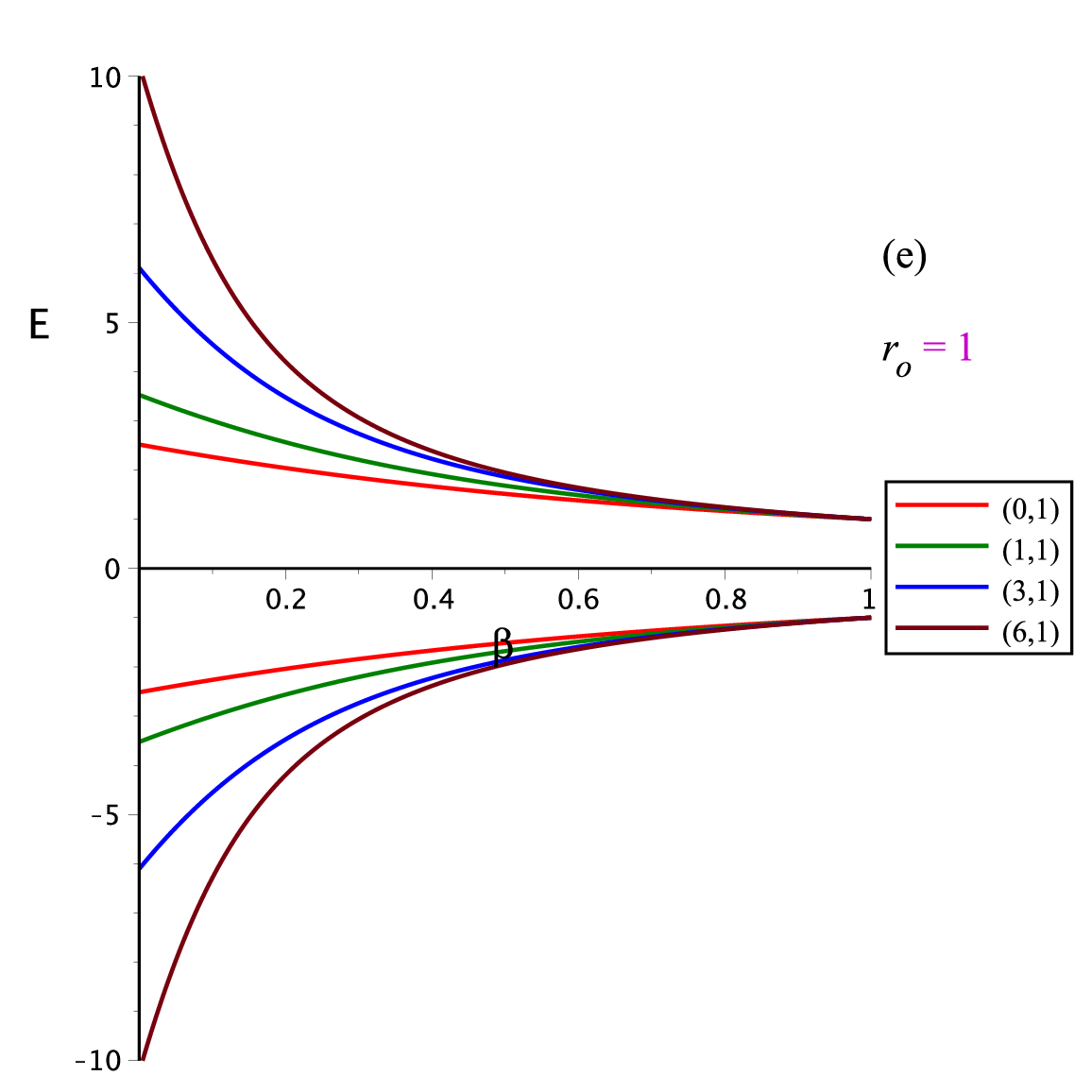}
\includegraphics[width=0.3\textwidth]{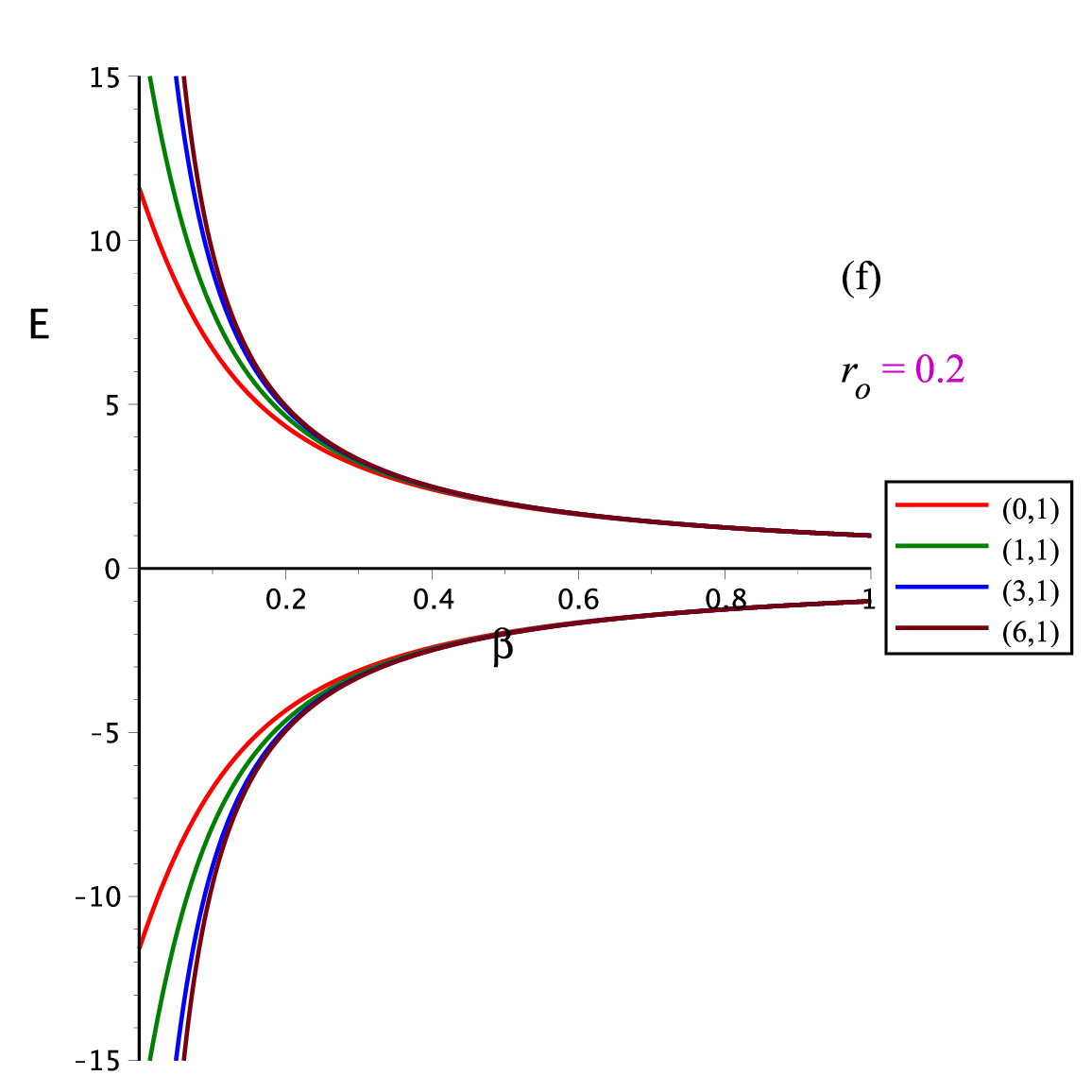}
\caption{\small 
{ For different $\left( n,m\right) $-states and $m_{0
}=m_{0 }c^{2}=1$, we show the KG-oscillator's energy levels in Eq.(\ref%
{34}) (a) against the oscillator's frequency $\Omega $ at $\alpha =0.5$ and $%
\beta =0$ (i.e., no rainbow gravity), (b) against the oscillator's frequency 
$\Omega $ at $\alpha =0.5$ and $\beta =0.5$, (c) against the TWH-throat
radius $r_{0 }$ at $\alpha =0.5$ and $\beta =0.5$, (d) against the
disclination parameter $\alpha $ at $\beta =0.5$ and $r_{0 }=1$, (e)
against the rainbow parameter $\beta =\epsilon /E_{P}$ at $\alpha =0.5$ and $%
r_{0 }=1$, and (f) against the rainbow parameter $\beta $ at $\alpha =0.5
$ and $r_{0 }=0.2$}}
\label{fig1}
\end{figure}
\begin{figure}[ht!]  
\centering
\includegraphics[width=0.3\textwidth]{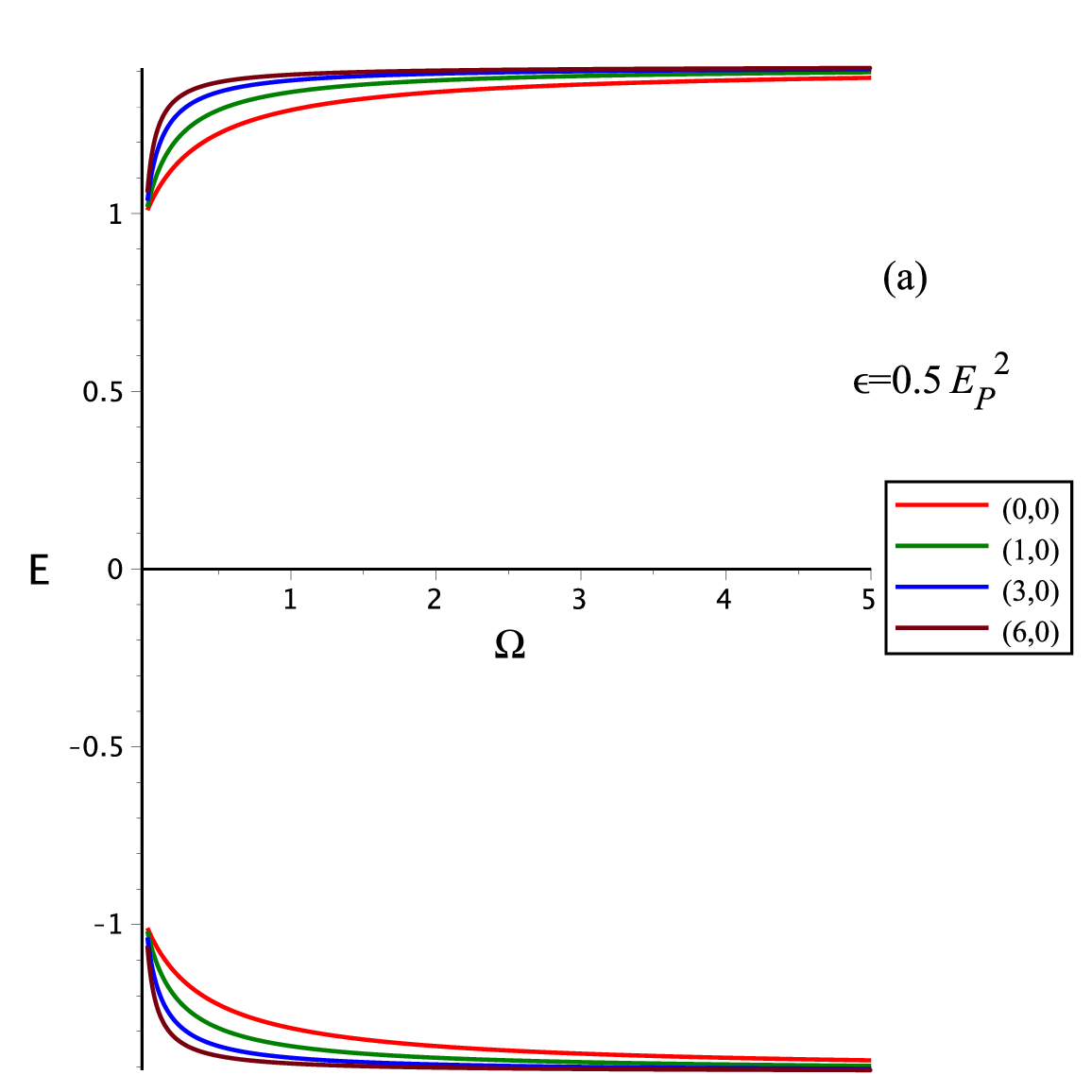}
\includegraphics[width=0.3\textwidth]{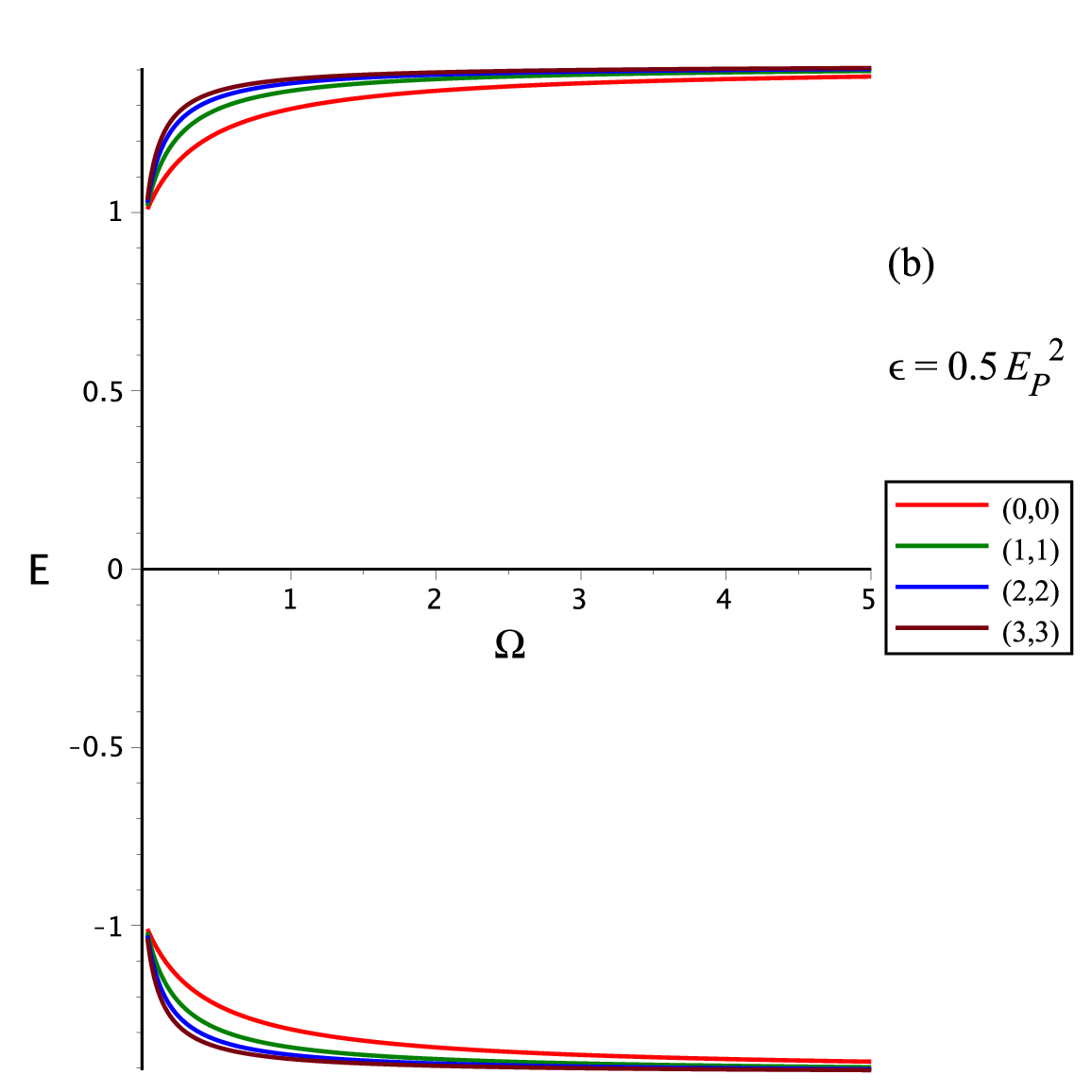}
\includegraphics[width=0.3\textwidth]{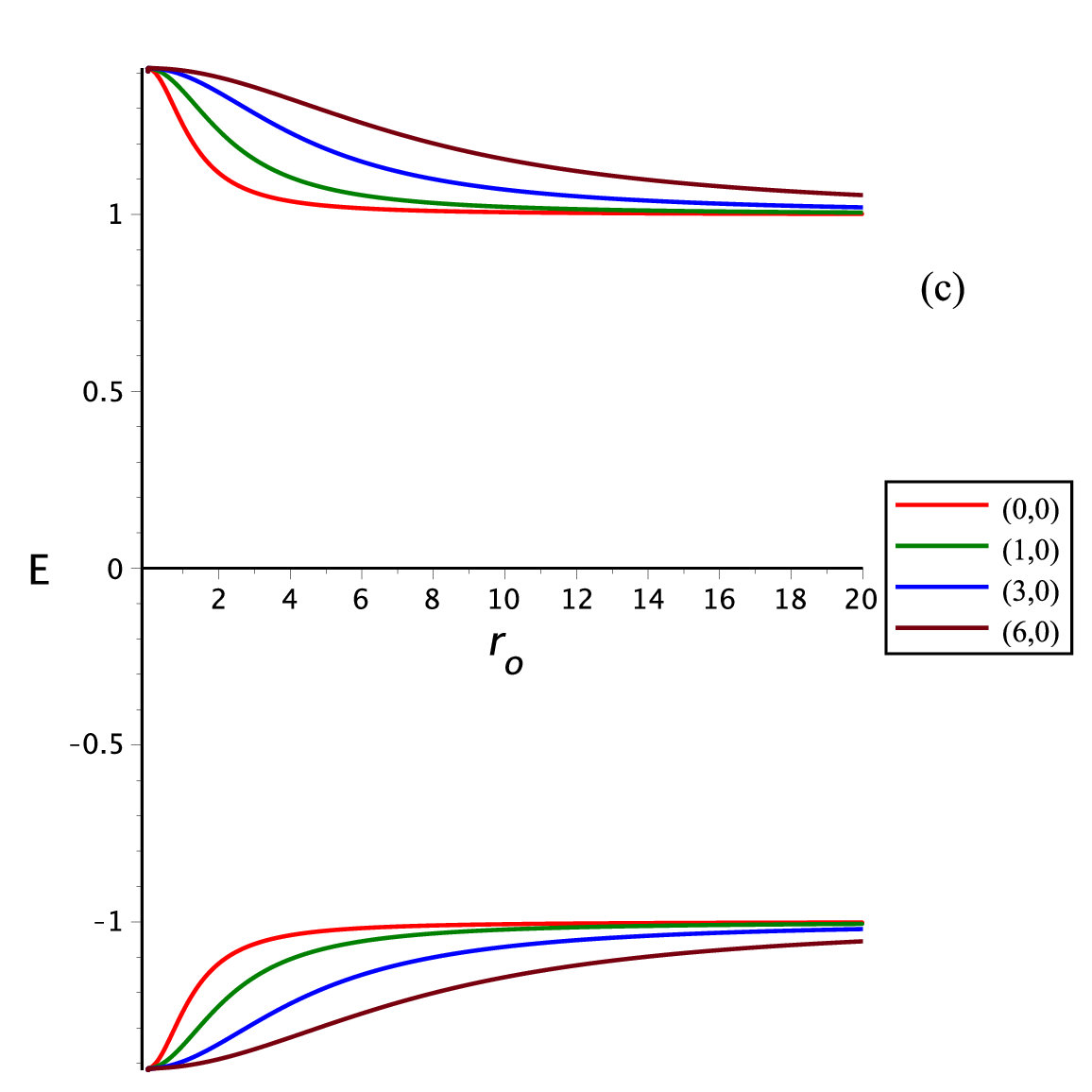}
\includegraphics[width=0.3\textwidth]{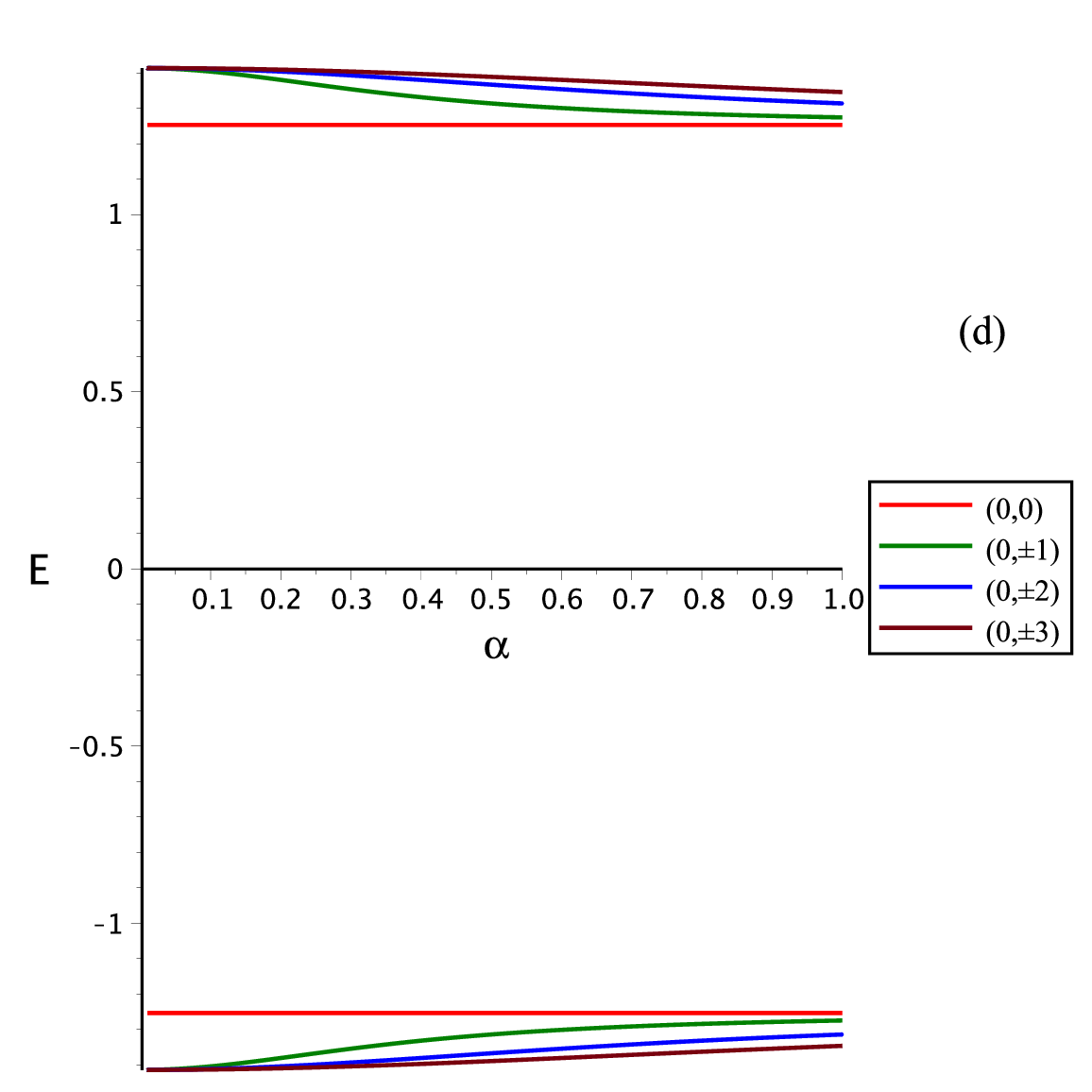}
\includegraphics[width=0.3\textwidth]{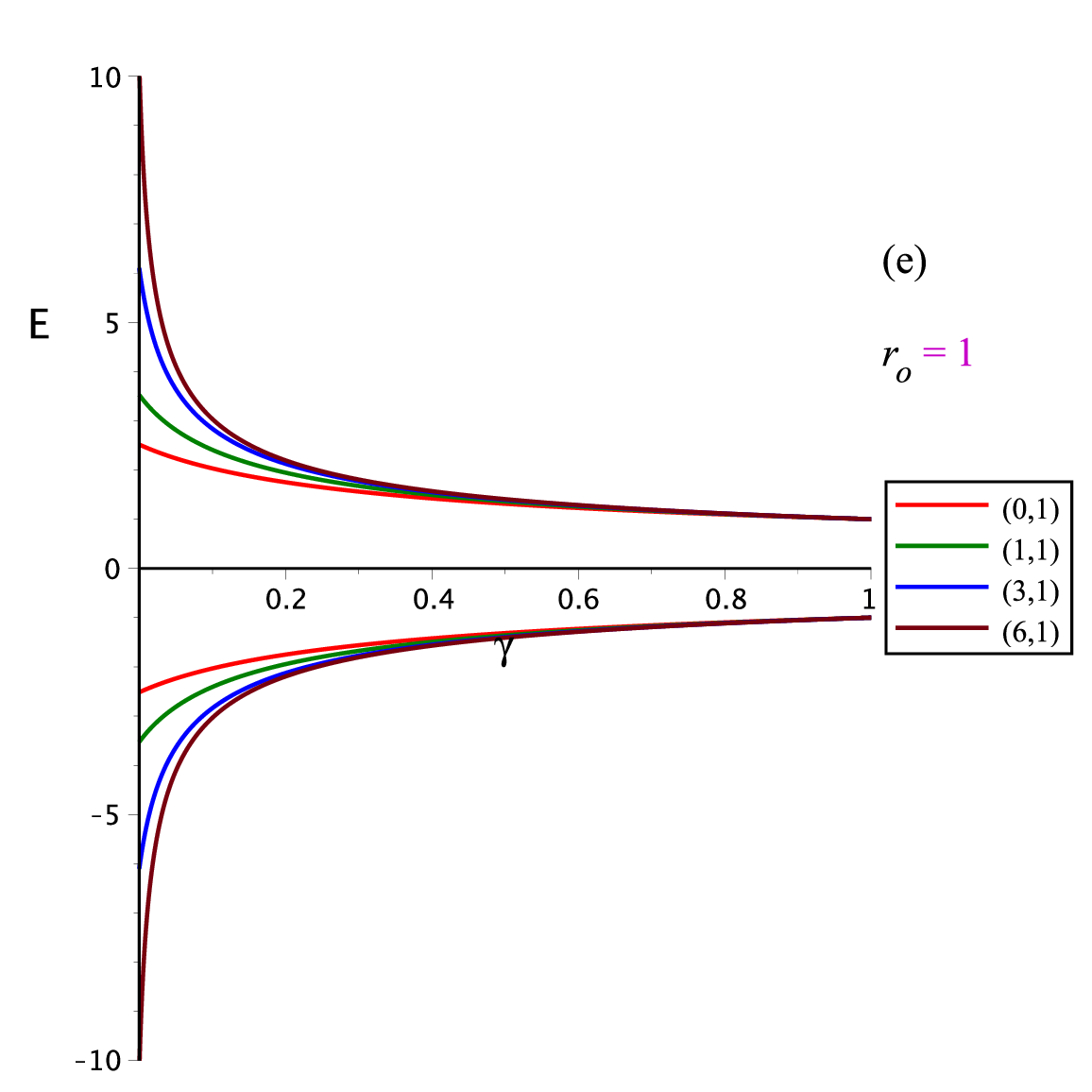}
\includegraphics[width=0.3\textwidth]{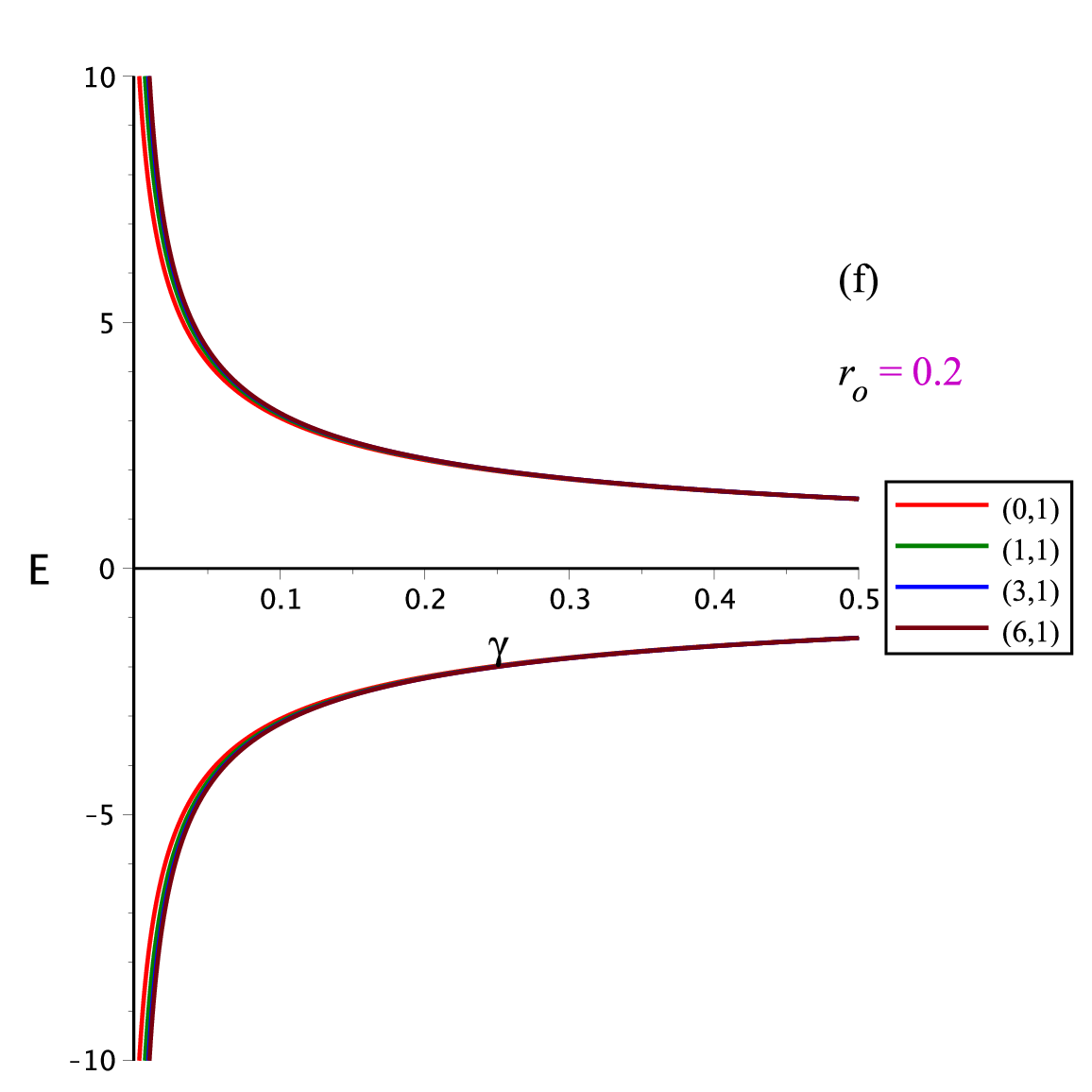}
\caption{\small 
{For different $\left( n,m\right) $-states and $m_{0
}=m_{0 }c^{2}=1$, we show the KG-oscillator's energy levels in Eq.(\ref
{29}) (a) against the oscillator's frequency $\Omega $ at $\alpha =0.5$, $m=0
$, and $\gamma =0.5$, (b) against the oscillator's frequency $\Omega $ at $
\alpha =0.5$, $\gamma =0.5$, and $m=0,\pm 1,\pm 2,\pm 3$, (c) against the
TWH-throat radius $r_{0 }$ at $\alpha =0.5$ and $\gamma =0.5$, (d)
against the disclination parameter $\alpha $ at $\gamma =0.5$ and $r_{0 }=1$
, (e) against the rainbow parameter $\gamma =\epsilon /E_{P}^{2}$ at $\alpha
=0.5$ and $r_{0 }=1$, and (f) against the rainbow parameter $\gamma $ at 
$\alpha =0.5$ and $r_{0 }=0.2$.}}
\label{fig1}
\end{figure}

In Figure 1 we plot the energy levels for different $\left( n,m\right) $-states (using $m_{0 }=m_{0 }c^{2}=1$). The asymptotic tendencies of the energies in (\ref{31}) and (\ref{33}) for $\Omega >>1$ are  $E_{\pm
}\sim \pm 1/\beta =\pm E_{P}/\epsilon \Longrightarrow \left\vert E_{\pm
}\right\vert \leq E_{P}$ (since the rainbow gravity parameter $\epsilon \geq
1$). Whereas, for $\Omega <<1$ the energies $E_{\pm }\sim \pm m_{0 }=\pm
m_{0 }c^{2}$ as documented in 1(a) and 1(b). Consequently, it is clear that the current loop quantum gravity motivated rainbow functions pair (i.e., $g_{_{0}}(u)=1$, $g_{_{1}}\left( u\right) =\sqrt{1-\epsilon u}
;\;u=|E|/E_{P}$), completely complies with the rainbow gravity model, and secures Planck's energy $E_{P}$ as the maximal energy for both particles and
anti-particles alike. This behavior may clearly be observed in Fig. 1(b) as our $E_{\pm }\sim \pm 1/\beta =\pm 2$ ( for $\beta =0.5$ value used in 1(b)) . Yet, a comparison between Fig. 1(a), without rainbow gravity, and 1(b), with rainbow gravity, one (undoubtedly) observes this rainbow gravity effect on the spectroscopic structure of the KG-particles and antiparticles in TWH-spacetime. Moreover, for the throat radius $r_{0 }<<1$ (using our
conditional exact solvability correlation (\ref{27})) the energies tend to approach $E_{\pm }\sim \pm 1/\beta =\pm E_{P}/\epsilon \Longrightarrow
\left\vert E_{\pm }\right\vert \leq E_{P}$ (i.e., $E_{\pm }\sim \pm 1/\beta
=\pm 2$  for $\beta =0.5$ value used in 1(c)). However, for  $r_{0
}>>\infty $ the energies tend to cluster around $E_{\pm }\sim \pm m_{0 }$.   This behavior is, in fact, an exact reflection of the inverse proportionality relation between the KG-oscillator frequency and the throat radius in (\ref{27}).  In 1(d) we observe that for the magnetic quantum number $m=0$, the disclination parameter $\alpha $ has no effects. However, as $\alpha $ grows up from just above zero the energy levels start splitting, but with degeneracies associated with $
m=\pm |m|$ (as a consequence of the correlation  (\ref{27})). Of course, one
would need a magnetic field to observe $\pm |m|$ splittings. When the rainbow gravity is switched off at $\beta =0$ (i.e., the rainbow parameter $\epsilon
=0$), the energies of the KG-oscillators in the TWH-spacetime background are
obtained as $E_{\pm }=\pm \sqrt{m_{0 }^{2}+4\Omega \left( n+1\right) }$ (provided that $\Omega $ is given by the correlation (\ref{27})). This is
made clear in 1(e) and 1(f), for the throat radius $r_{0 }=1$ and $%
r_{0 }=0.2$, respectively. For example, for $\beta =0$, $n=0$ and $m=0$, we have $\Omega =100/3$ and therefore $E_{\pm }=\pm 11.59$ as shown in 1(f). The effect of the throat radius $r_{0 }$, on the other hand, is identified through the comparison between 1(e) and 1(f). We clearly observe
that the energy levels tend to cluster more rapidly as the throat radius $
r_{0 }$ decreases. 

\subsection{Rainbow function pair $g_{_{0}}(u)=1$, $g_{_{1}}\left( u\right) =
\sqrt{1-\epsilon u^{2}}$}

In this case, Eq.(\ref{28}) would result, with $\gamma =\epsilon /E_{P}^{2}$%
, that%
\begin{equation}
E^{2}=m_{0 }^{2}+4\Omega \left( n+1\right) \left( 1-\gamma E^{2}\right)
\Longrightarrow E=\pm \sqrt{\frac{m_{0 }^{2}+4\Omega \left( n+1\right) }{%
1+4\Omega \left( n+1\right) \gamma }}.  \label{34}
\end{equation}%
Where the asymptotic tendency, as  $\Omega >>1$ is $E_{\pm }\sim \pm 1/\sqrt{
\gamma }=\pm E_{P}/\sqrt{\epsilon }\Longrightarrow \left\vert E_{\pm
}\right\vert \leq E_{P}$ (for the rainbow parameter $\epsilon \geq 1$).
Whereas, as $\Omega <<1$ we, again, observe that this loop quantum gravity
motivated rainbow functions pair (i.e., $g_{_{0}}(u)=1$, $g_{_{1}}\left(
u\right) =\sqrt{1-\epsilon u^{2}}$) completely complies with the rainbow
gravity model. This behavior may clearly be observed in Fig. 2(a) and 2(b)
as our $E_{\pm }\sim \pm 1/\sqrt{\gamma }=\pm 1.4$ ( for $\gamma =0.5$ value
used in both figures). For the throat radius $r_{0 }<<1$ the energies
tend to approach $E_{\pm }\sim \pm 1/\sqrt{\gamma }=\pm E_{P}/\sqrt{\epsilon 
}\Longrightarrow \left\vert E_{\pm }\right\vert \leq E_{P}$ (i.e., $E_{\pm
}\sim \pm 1/\sqrt{\gamma }=\pm 1.4$ for $\gamma =0.5$ value used in 2(c)). However, for $r_{0 }>>\infty $ the energies tend to cluster around $
E_{\pm }\sim \pm m_{0 }$. In 2(d) we observe that as $\alpha $ grows up from just above zero the energy levels start splitting.but with degeneracies
associated with $m=\pm |m|$ as a consequence of the correlation (\ref{27}).
The comparison between 1(e) and 1(f) shows that the energies cluster more
rapidly as $r_{0 }$ decreases. 

\section{Concluding remarks}

In the current proposal, we have investigated KG-oscillators in TWH-spacetime background embedded with disclination ($\propto \alpha$) in rainbow gravity. We have started with a discussion, in short,
on the KG-particles using the non-minimal coupling form $\tilde{D}^\pm
_{i}=\partial_{i}\pm\mathcal{F}_{i}$, where  $\mathcal{F}_{i}=\left( 0,\mathcal{F}_{x},0\right) $.  Next, we have used Moshinsky-Szczepaniak's \cite{RR3} and Mirza-Mohadesi's \cite{RR4} recipes, i.e.,  $\mathcal{F}_{x}=\Omega x
$, to incorporate KG-oscillators. However, we have replaced their $m_{0 }\omega $ by our $
\Omega $ so that the reader does not confuse $m_{0 }$ (the particle's mass in Schr\"{o}dinger-oscillators problem) with $m_{0 }=m_{0 }c^{2}
$  (the rest mass energy of the KG-particle) on the RHS of the KG-equation in (\ref{6}), they are simply not the same.
Under such settings, we have discussed a conditional exact solution (very recently developed and used in e.g., \cite{RR1,RR2,RR2.1,RR2.2,RR2.3,RR2.4,RR2.5} for the corresponding KG-oscillators and show that the truncation of the confluent Heun function is truncated into a polynomial of order $n+1$ is associated with the parametric correlation (\ref{27}). Such a conditional exact solvability of the KG-oscillators in TWH-spacetime with and without rainbow gravity have provided a vast number of $\left( n,m\right) $-states to be
investigated as to the effects of all parameters. In the process, we have
used two loop quantum gravity motivated pairs $g_{_{0}}(u)=1$, $
g_{_{1}}\left( u\right) =\sqrt{1-\epsilon u^{2}}$, and $g_{_{0}}(u)=1$, $
g_{_{1}}\left( u\right) =\sqrt{1-\epsilon u}$, where $\epsilon \geq 1$ is
the rainbow gravity parameter \cite{R15,R15.1,R16,R17}.

In the light of our experience above, our observations are in order. In
connection with the KG-oscillators' frequency $\Omega $, we have observed
that when $\Omega >>1$ all energy levels cluster around the value $E_{\pm
}\sim \pm 1/\beta =\pm E_{P}/\epsilon \Longrightarrow \left\vert E_{\pm
}\right\vert \leq E_{P}$ for $g_{_{1}}\left( u\right) =\sqrt{1-\epsilon u}$,
and $E_{\pm }\sim \pm 1/\sqrt{\gamma }=\pm E_{P}/\sqrt{\epsilon }
\Longrightarrow \left\vert E_{\pm }\right\vert \leq E_{P}$ for $
g_{_{1}}\left( u\right) =\sqrt{1-\epsilon u^{2}}$. This is documented in
figures 1(b) and 2(a)/2(b), respectively.  These observations agree with the
asymptotic tendencies of the results in (\ref{31}), (\ref{33}), and (\ref{34}) when $\Omega >>1$. For the TWH throat 
radius $r_{0 }$, on the
other hand, we have observed that for $r_{0 }>>\infty $ the energies
tend to cluster around $E_{\pm }\sim \pm m_{0 }$ (indicating that the
KG-particles admit no bound states near the asymptotically flat upper and
lower universes connected by the TWH), whereas for $r_{0 }<<1$ the
energies tend to approach $E_{\pm }\sim \pm 1/\beta =\pm E_{P}/\epsilon
\Longrightarrow \left\vert E_{\pm }\right\vert \leq E_{P}$. This is
documented in figures 1(c) and 2(c). Moreover, in connection with the effect
of both the rainbow parameter and the TWH throat radius, we have observed
that while increasing the rainbow gravity parameter ($\beta =\epsilon /$ $
E_{P}\geq 0$ in the case of $g_{_{1}}\left( u\right) =\sqrt{1-\epsilon u}$
or $\gamma =\epsilon /E_{P}^{2}\geq 0$ in the case of $g_{_{1}}\left(
u\right) =\sqrt{1-\epsilon u^{2}}$), the energy levels tend to cluster more
rapidly as the throat radius $r_{0 }$ decreases. This effect is made
obvious in figures 1(e), 1(f) 2(e), and 2(f).

Finally, even with our \textit{conditionally exact solvability}, through the
correlation (\ref{27}), of the KG-oscillators in TWH-spacetime in rainbow
gravity, we have had a chance to comprehensively discuss the effects of the
TWH throat radius along with the rainbow gravity parameter on the
spectroscopic structure of the KG-oscillators. Therefore, the current
methodical proposal, in effect, provides some detailed/comprehensive 
analysis for a set of KG-oscillators bound states in TWH spacetime in
rainbow gravity. To the best of our knowledge, such results have never been
reported elsewhere and are therefore new results.


\section*{\small{Data availability}}
This manuscript has no associated data or the data will not be deposited.

\section*{\small{Conflicts of interest statement}}
No conflict of interest declared by the authors.

\section*{\small{Funding}}
No funding regarding this research.

\end{document}